\documentstyle[12pt]{article}

\font\titlefont=cmbx10 scaled \magstep3
\begin{document}
\input{epsf}

\begin{flushright}
\vspace*{-2cm}
  gr-qc/9704050  \\ 
April 17, 1997 \\
CBPF-NF-017/97  \\
TUTP-97-8
\vspace*{0.5cm}
\end{flushright}

\begin{center}
{\titlefont COSMOLOGICAL AND BLACK HOLE \\
\vskip 0.15in
HORIZON FLUCTUATIONS }
\vskip .3in
L.H. Ford\footnote{email: ford@cosmos2.phy.tufts.edu} \\
\vskip .1in
Institute of Cosmology,
Department of Physics and Astronomy\\
Tufts University\\
Medford, Massachusetts 02155\\
\vskip .2in
N.F. Svaiter\footnote{email: nfuxsvai@lca1.drp.cbpf.br} \\ 
\vskip .1in
Centro Brasileiro de Pesquisas Fisicas-CBPF \\ 
Rua Dr. Xavier Sigaud 150\\ 
Rio de Janeiro, RJ 22290-180, Brazil \\
\end{center}

\vskip .2in

\begin{abstract}
The quantum fluctuations of horizons in Robertson-Walker universes and
in the Schwarzschild spacetime are discussed. The source of the metric 
fluctuations is taken to be quantum linear perturbations of the gravitational
field. Lightcone fluctuations arise when the retarded Green's function
for a massless
field is averaged over these metric fluctuations. This averaging replaces the 
delta-function on the classical lightcone with a Gaussian function, the width
of which is a measure of the scale of the lightcone fluctuations. Horizon
fluctuations are taken to be measured in the frame of a geodesic observer
falling through the horizon. In the case of an expanding universe, this is
a comoving observer either entering or leaving the horizon of another 
observer. In the black hole case, we take this observer to be one who falls 
freely from rest at infinity. We find that cosmological horizon fluctuations
are typically characterized by the Planck length. However, black hole horizon
fluctuations in this model are much smaller than Planck dimensions for black
holes whose mass exceeds the Planck mass. Furthermore, we find black hole 
horizon fluctuations which are sufficiently small as not to invalidate the 
semiclassical derivation of the Hawking process.
\end{abstract}
\vskip .1in
{\bf PACS numbers:} 04.60.-m, 04.70.Bw, 04.62.+v
\newpage

\baselineskip =14pt

\section{Introduction}
\label{sec:Intro}

      One of the characteristics of classical gravitation is the existence of
horizons, surfaces which divide spacetime into causally distinct regions.
The most striking example is the black hole horizon, the boundary which hides
the events within from the outside world. Cosmological models also possess
horizons of a different sort; a given observer generally cannot see all of the 
other observers in the universe at a given time. If the expansion rate in
comoving time is less than linear, then previously unseen objects enter
the observer's horizon. If it is faster than linear (inflationary expansion), 
then objects leave the horizon. Horizons are of course lightcones, and
the notion of an event being within or without a horizon means being at a
timelike or a spacelike separation, respectively.

     It is expected that quantum metric fluctuations should smear out this
precise distinction, and hence smear out the classical concept of a horizon.
Information could presumably leak across the horizon in a way that is not
allowed by classical physics. Bekenstein and Mukhanov \cite{BM95} have 
suggested that horizon fluctuations could lead to discreteness of the spectrum
of black holes.
Several other authors \cite{Sorkin,Casher} have  
recently made proposals for models which describe the horizon fluctuations.
In this paper, we will propose a different model, in which quantized linear
perturbations of the gravitational field act as the source of the underlying
metric fluctuations. Our analysis will be based on the formalism for the
study of lightcone fluctuations proposed in Ref. \cite{F95}, and further
developed in Ref. \cite{FS96}. 

      The necessary formalism will be reviewed and extended in 
Sect.~\ref{sec:Basic}. It will be applied to the case of cosmological horizons 
in Sect.~\ref{sec:Cosmo}, and to black hole horizons in Sect.~\ref{sec:BH}.   
Our results will be summarized and discussed in Sect.~\ref{sec:Sum}. We will
also give a critical assessment of the previous attempts \cite{Sorkin,Casher}
to estimate the horizon fluctuations.

\section{Basic Formalism}
\label{sec:Basic}

     In Ref. \cite{F95}, henceforth I, a model of lightcone fluctuations
on a flat background was developed. It was assumed that the quantized
gravitational field is in a squeezed vacuum state. This is the natural
quantum state for gravitons produced by quantum particle creation
processes, as for example in the early universe. Here we wish to
generalize this formalism to the case of curved background spacetimes.
Consider an arbitrary background metric $g_{\mu\nu}^{(0)}$ with a linear
perturbation $h_{\mu\nu}$, so the spacetime metric is \cite{units}
\begin{equation}
ds^2 = (g_{\mu\nu}^{(0)} + h_{\mu\nu}) d x^\mu d x^\nu \,. \label{eq:metric}
\end{equation}
For any pair of spacetime points $x$ and $x'$, let $\sigma(x,x')$ be one half
of the squared geodesic separation in the full metric, and $\sigma_0(x,x')$ 
be the corresponding quantity in the background metric. We can expand
$\sigma(x,x')$ in powers of $h_{\mu\nu}$ as
\begin{equation}
\sigma = \sigma_0 + \sigma_1 + \sigma_2 + \cdots \, ,
\end{equation}
where $\sigma_1$ is first order in $h_{\mu\nu}$, ect. We now suppose
that the linearized perturbation $h_{\mu\nu}$ is quantized, and that the
quantum state $|\psi \rangle$ is a ``vacuum'' state in the sense that
we can decompose $h_{\mu\nu}$ into positive and negative frequency parts
$h^{+}_{\mu\nu}$ and $h^{-}_{\mu\nu}$, respectively, such that
\begin{equation}
h^{+}_{\mu\nu} |\psi \rangle =0, \qquad \langle \psi|h^{-}_{\mu\nu} =0 \,.
\end{equation}
It follows immediately that
\begin{equation}
\langle  h_{\mu\nu} \rangle =0
\end{equation}
in state $|\psi \rangle$. In general, however, 
$\langle (h_{\mu\nu})^2 \rangle \not= 0$, where the expectation value is
understood to be suitably renormalized. This reflects the quantum metric
fluctuations.

    We now wish to average the retarded Green's function, $G_{ret}(x,x')$,
for a massless field over the metric fluctuations. In a curved
spacetime, $G_{ret}(x,x')$ can be nonzero inside the future lightcone as
a result of backscattering off of the spacetime curvature. However, its
asymptotic form near the lightcone is the same as in flat spacetime:
\begin{equation}
G_{ret}(x,x') \sim {{\theta(t-t')}\over {4\pi}} \delta(\sigma) \, ,
        \quad  \sigma \rightarrow 0 \, .
                            \label{eq:gret0}
\end{equation}
We will ignore the backscattered portion, and average this
delta-function term over the fluctuations, following the method of I. The
result
is
\begin{equation}
\Bigl\langle G_{ret}(x,x') \Bigr\rangle = 
{{\theta(t-t')}\over {8\pi^2}} \sqrt{\pi \over {2\langle \sigma_1^2 \rangle}}
\; \exp\biggl(-{{\sigma_0^2}\over {2\langle \sigma_1^2 \rangle}}\biggr)\, .
                                           \label{eq:retav}
\end{equation}
The effect of the averaging has been to replace the delta-function by a
Gaussian with a finite width determined by the magnitude of the quantity
$\langle \sigma_1^2 \rangle$, which is the measure of the lightcone
fluctuations. 

     The operational meaning of the smeared lightcone can be understood
by considering a source and a detector of photons. If we ignore the
finite sizes of photon wavepackets, then in the absence of lightcone
fluctuations, all photons should traverse the interval between the
source and the detector in the same amount of time. The effect of the
lightcone fluctuations is to cause some photons to travel slower than
the classical light speed, and others to travel faster. The Gaussian 
function in Eq.~(\ref{eq:retav}) is symmetrical about the classical lightcone,
$\sigma_0 = 0$, so the quantum lightcone fluctuations are equally likely
to produce a time advance as a time delay. 

    In order to find the magnitude of the lightcone fluctuations in a
particular situation, it is necessary to calculate $\sigma_0$ for the
metric in question, as well as $\langle \sigma_1^2 \rangle$ in the
appropriate quantum state. This enables one to find $\Delta t$, the mean time
delay or advance (measured in a suitable reference frame). This is an
ensemble averaged quantity, not necessarily the expected variation in flight
time of two photons emitted in rapid succession. To find the latter
quantity, one must examine a correlation function. This is the topic of
Ref. \cite{FS96}. In the present paper, we will not be concerned with 
correlation functions, and will use Eq.~(\ref{eq:retav}) to estimate the
magnitude of the horizon fluctuations. 

     We may find a general expression for $\langle \sigma_1^2 \rangle$,
which is the curved space generalization of the result obtained in I.
Let us first consider timelike geodesics. If we adopt a timelike metric,
then in Eq.~(\ref{eq:metric}) we have that $d s^2 > 0$. Let 
$u^\mu = dx^\mu /d\tau$ be the tangent to the geodesic and $\tau$ be the 
proper time. We will define $\langle \sigma_1^2 \rangle$ by integrating
along the unperturbed geodesic, in which case $u^\mu$ is normalized to unity in 
the background metric:
\begin{equation}
g_{\mu\nu}^{(0)} u^\mu u^\nu = 1\,.
\end{equation}
The geodesic interval in the unperturbed metric is given by
\begin{equation}
\sigma_0 = \frac{1}{2} (\Delta \tau)^2 \,,
\end{equation}
where $\Delta \tau$ is the proper time elapsed along the geodesic.
We have
\begin{equation}
\frac{d s}{d\tau} = \sqrt{ 1+ h_{\mu\nu}\, u^\mu u^\nu}
                  \approx 1 + \frac{1}{2} h_{\mu\nu}\, u^\mu u^\nu \,,
\end{equation}
and hence the geodesic length between a pair of points in the perturbed
metric is $\Delta s = \Delta \tau + \Delta s_1$, where
\begin{equation}
\Delta s_1 = \frac{1}{2} \int d\tau\, h_{\mu\nu}\, u^\mu u^\nu \,.
\end{equation}
Thus
\begin{equation}
\sigma = \frac{1}{2} (\Delta s)^2 = \frac{1}{2} (\Delta \tau)^2 
        + \Delta \tau \Delta s_1 + O(h^2) \,,
\end{equation}
and hence $\sigma_1 = \Delta \tau \Delta s_1$. If we average over the 
metric fluctuations, the result is
\begin{equation}
\langle \sigma_1^2 \rangle = \frac{1}{2} \sigma_0 
\int d\tau_1\, d\tau_2\, u_1^\mu u_1^\nu u_2^\rho u_2^\sigma \,
                     \langle h_{\mu\nu}(x_1) h_{\rho\sigma}(x_2) \rangle \,,
                                 \label{eq:sig1gen}
\end{equation}
where $u_1^\mu = dx^\mu /d\tau_1$ and $u_2^\mu = dx^\mu /d\tau_2$. 
An analogous expression holds for the case of a spacelike geodesic, in which
the integrations are over the proper length parameter of the geodesic:
\begin{equation}
\langle \sigma_1^2 \rangle = - \frac{1}{2} \sigma_0 
\int d\lambda_1\, d\lambda_2\, u_1^\mu u_1^\nu u_2^\rho u_2^\sigma \,
                     \langle h_{\mu\nu}(x_1) h_{\rho\sigma}(x_2) \rangle \,,
                                 \label{eq:sig1gen2}
\end{equation}
where now $u_1^\mu = dx^\mu /d\lambda_1$ is the tangent to the geodesic, and
$\lambda$ is the proper length. Here we have $\sigma_0 = 
 - \frac{1}{2} (\Delta \lambda)^2$.

   As noted previously, the quantity $\langle \sigma_1^2 \rangle$ is formally
divergent and needs to be renormalized. This may be done by defining the
graviton two-point function, 
$\langle h_{\mu\nu}(x_1) h_{\rho\sigma}(x_2) \rangle$ using, for example, 
the Hadamard renormalization scheme proposed by
Brown and Ottewill \cite{BO86}. These authors give a detailed prescription for
expanding the singular, state-independent parts of the scalar and vector
two-point functions in an arbitrary curved spacetime. Hadamard renormalization
consists of subtracting this expansion from a given two-point function. 
This procedure seems not to have been developed in detail for the graviton
two-point function, but there seems to be no barrier in principle to doing so.
Allen {\it et al} \cite{AMO92} have applied the Hadamard renormalization
method to the graviton two-point function in the vicinity of a cosmic string.
In this paper, we will be content with simple approximations or order of magnitude 
estimates, and will not require the full renormalization machinery.

\section{Cosmological Horizons}
\label{sec:Cosmo}

    Consider a spatially flat Robertson-Walker universe, for which the metric 
may be written as
\begin{equation}
ds^2 = a^2(\eta) (d\eta^2 - d{\bf x}^2 ) \,. \label{eq:RWmetric}
\end{equation}
In general, this spacetime has ``particle horizons'' associated with the 
comoving observers across which other observers may appear or disappear.
In the case of a radiation or matter dominated universe with an initial
singularity ($a \propto \eta$ or $a \propto \eta^2$, respectively), a
given observer at a given time has not yet received any light signals from
distant observers, who are said to be outside of the first observer's
horizon. In the case of de Sitter space ($a \propto \eta^{-1}$), a given
observer eventually ceases to receive signals from other comoving observers,
and views them as having moved outside of the horizon. Clearly, these
cosmological horizons are observer dependent in a way that black hole event 
horizons are not, and are basically the past lightcone of a given observer at 
a given time. Nonetheless, it will be of interest to estimate the magnitude
of the quantum fluctuation of these horizons in various models. 

    We must first study timelike and spacelike geodesics in the limits in
which these approach null geodesics over some interval. Because the lightcone
fluctuations are symmetrical, we may focus our attention on the timelike case.
The geodesic equations for a timelike observer moving in the $x$-direction in 
the metric of Eq.~(\ref{eq:RWmetric}) may be expressed as 
\begin{equation}
a^2 \frac{d x}{ d\tau} = \frac{1}{\sqrt{2\alpha}}  \label{eq:geod1}
\end{equation}
and 
\begin{equation}
\frac{d^2 \eta}{d \tau^2} + \frac{a'}{a} \biggl(\frac{d \eta}{d \tau}\biggr)
 + \frac{a'}{2\alpha a^5} =0 \, , \label{eq:geod2}
\end{equation}
where $\tau$ is the proper time along the geodesic, $a' =d a/d \eta$, and
$\alpha$ is a constant. In the flat space limit ($a=1$), we find from
Eq.~(\ref{eq:geod1}) that $\alpha = (1-v^2)/(2v^2)$ where $v$ is the magnitude
of the three-velocity. Thus in the null limit, $\alpha \rightarrow 0$.
The geodesic for a particle which starts at $\eta = \eta_0$ at $x = 0 $ 
may be expressed as 
\begin{equation}
x = \eta -\eta_0 - f(\eta,\eta_0)\, ,   \label{eq:def_f}
\end{equation}
where $f \rightarrow 0$ in the null limit. For nearly null geodesics, we may 
assume $|f| \ll 1$. To first order in $f$, the solution of 
Eqs.~(\ref{eq:geod1}) and (\ref{eq:geod2}), corresponding to
a properly normalized four-velocity, is
\begin{equation}
f(\eta,\eta_0) = \alpha \int_{\eta_0}^\eta a^2(\eta') d \eta' \, .  
                                       \label{eq:f}
\end{equation}
From Eq.~(\ref{eq:RWmetric}), we can write
\begin{equation}
d s = \sqrt{ 1-\biggl(\frac{dx}{d \eta}\biggr)^2}\; a \, d\eta
                  \approx \sqrt{2\alpha}\, a^2 \, d\eta\,,
\end{equation}
from which we find
\begin{equation}
\Delta \tau = \sqrt{2\alpha} \int_{\eta_0}^{\eta_1} a^2(\eta) d \eta \,,
                                               \label{eq:s0}
\end{equation}
and 
\begin{equation}
\sigma_0 = \frac{1}{2} (\Delta \tau )^2 \,.
\end{equation}

     Gravitons propagating on a Robertson-Walker background may be quantized
in the transverse, trace-free gauge, which eliminates all of the gauge 
freedom \cite{FP77}. Only the purely spatial components of $h^\mu_\nu$ are
nonzero, and they each satisfy the wave equation for a massless, minimally
coupled scalar field in this metric. Thus the graviton two-point function
can be expressed in terms of the scalar two-point function, 
$\langle \varphi(x) \varphi(x') \rangle$, as
\begin{equation}
\langle h_{ij(x)}h_{kl}(x') \rangle = 
-\frac{1}{3} a^2(\eta) a^2(\eta') \Bigl(\delta_{ij}\delta_{kl} -
{3\over 2}\delta_{ik}\delta_{jl} -{3\over 2}\delta_{il}\delta_{jk} \Bigr)
\langle \varphi(x) \varphi(x') \rangle \,.
\end{equation}
We may use this result to write
\begin{equation}
\langle \sigma_1^2 \rangle = \frac{\sigma_0}{6\alpha}  
\int_{\eta_0}^{\eta_1} d\eta \int_{\eta_0}^{\eta_1} d\eta'
          \,   \langle \varphi(x) \varphi(x') \rangle \,,  \label{eq:sig1}
\end{equation}
Note that in this expression, we can replace the unsymmetrized two point
function, $ \langle \varphi(x) \varphi(x') \rangle$, by the symmetrized
form (the Hadamard function)
\begin{equation}
G(x,x') = \frac{1}{2} \langle \varphi(x) \varphi(x') +
          \varphi(x')  \varphi(x) \rangle \,.
\end{equation}
The latter function is real, and is hence more convenient.
To proceed further, we must designate the quantum state of the gravitons.
In the following two subsections, some particular examples will be considered.

\subsection{Gravitons in a Radiation-Dominated Universe}

     A radiation-dominated universe, for which
\begin{equation}
a(\eta) = a_0 \, \eta \, , \label{eq:raddom}
\end{equation}
is presumably a reasonably good description of a significant fraction of the
history of our universe. Let us consider a thermal bath of gravitons in
such a universe, for which the temperature is always high compared to the
scale set by the local radius of curvature, i.e., the thermal wavelength
is much less than the horizon size. In this case, the minimally coupled
scalar field two point function is approximately equal to that for the 
conformally coupled field, $G_{cc}(x,x')$. However, the latter is conformally 
related to the flat space Hadamard function, $G_0(x,x')$, so we have
\begin{equation}
G(x,x')  \approx G_{cc}(x,x')   =
   a^{-1}(\eta)\, a^{-1}(\eta')\,G_0(x,x')  \,.
\end{equation}
We now need the flat space renormalized thermal Green's function on the 
lightcone. This was calculated in Appendix A of Ref. \cite{FS96}, where it
was shown that in the high temperature limit, this function is given by
\begin{equation}
G_0(x,x') \approx \frac{1}{8\pi \beta \rho}
       \qquad \beta \ll \rho   \,,  \label{eq:highT}
\end{equation}
where $\beta$ is the inverse temperature and $\rho =|{\bf x} - {\bf x'}|$.
We may now use Eqs.~(\ref{eq:sig1}) - (\ref{eq:highT}) to write
\begin{equation}      
\langle \sigma_1^2 \rangle = \frac{\sigma_0}{48 \pi\,\alpha\,\beta\, a_0^2}
\int_{\eta_0}^{\eta_1} d\eta \int_{\eta_0}^{\eta_1} d\eta' 
  \frac{1}{\eta \eta'|\eta -\eta'|} \,.
\end{equation}
Unfortunately, this integral diverges because of the singularity of the
integrand at $\eta =\eta'$. This is due to the fact that Eq.~(\ref{eq:highT})
is not valid for small $\rho$. We can remedy this by excluding the range
$|\eta -\eta'| < \epsilon$, where $\epsilon$ is a cutoff which will be taken 
to be of order $\beta$. Thus, the relevant integral is, in the limit of
$\eta_1 \gg \eta_0$;
\begin{equation}
\langle \sigma_1^2 \rangle = \frac{\sigma_0}{48 \pi\,\alpha\,\beta\, a_0^2}
\int_{\eta_0}^{\eta_1} d\eta \;
\biggl(\int_{\eta_0}^\epsilon + \int_\epsilon^{\eta_1} \biggr) d\eta' \;
\frac{1}{\eta \eta'|\eta -\eta'|} \approx
\frac{\sigma_0 \Bigl[\ln(\eta_0/\epsilon) + 1 \Bigr]}
              {24\pi\,\alpha\,\beta\,\eta_0 a_0^2} \,. \label{eq:sig1a}
\end{equation}
In this same limit, one finds
\begin{equation}
\sigma_0 \approx \frac{1}{9} \alpha\, a^2_0\, \eta_1^6 \,. \label{eq:sig0a}
\end{equation}

   We now wish to define $\alpha_c$ as that value of $\alpha$ for which 
the argument of the exponential in Eq.~(\ref{eq:retav})  is unity, that is,
\begin{equation}
{{\sigma_0^2}\over {2\langle \sigma_1^2 \rangle}} =1 \, .
                                           \label{eq:alphac}
\end{equation}
Thus $\alpha_c$ describes a geodesic whose deviation from the lightcone
characterizes the fluctuations. From Eqs.~(\ref{eq:sig1a}), (\ref{eq:sig0a}),
and (\ref{eq:alphac}), we find that 
\begin{equation}
\alpha_c = \frac{\sqrt{3 \Bigl[\ln(\eta_0/\epsilon) + 1 \Bigr]}}
                {\sqrt{4\pi\alpha\beta\eta_0} \;a^3_0 \,\eta_1^3} \,.
\end{equation}
We need to define a physical measure of the magnitude of the lightcone
fluctuations. This may be taken to be the mean time delay or advance, 
$\Delta \eta$, for a photon emitted at $\eta_0$ and detected at $\eta_1$. 
Equivalently, we can think of $\Delta \eta$ as the characteristic interval
around $\eta_0$ within which photons could be emitted and still reach a
detector at a coordinate distance of $\Delta x = \eta_1 - \eta_0$ at time
$\eta_1$. (See Figure 1.) From Eqs.~(\ref{eq:def_f}) and (\ref{eq:f}), we
have that $\Delta \eta$ is related to $\alpha_c$ by
\begin{equation}
\Delta \eta = f(\eta_1,\eta_0) = 
\alpha_c \int_{\eta_0}^{\eta_1} a^2(\eta') d \eta' \approx 
  \frac{1}{3}\, \alpha_c\, a^2_0\, \eta_1^3 \, .  \label{eq:deleta}
\end{equation}
It is perhaps more convenient to express this time delay or advance as a
coordinate time interval, $\Delta t = a(\eta_0)\, \Delta \eta$, which is
given by
\begin{equation}
\Delta t = \biggl( \frac{2 t_0}{a_0}\biggr)^{\frac{1}{4}} \;
 \frac{\sqrt{3 \Bigl[\ln(\eta_0/\epsilon) + 1 \Bigr]}}{6 \sqrt{\pi\beta}} \,,
\end{equation}
where $t_0 = \frac{1}{2} a_0 \eta_0^2$ is the coordinate time at $\eta_0$.

\begin{figure}
\begin{center}
\leavevmode\epsfysize=8cm\epsffile{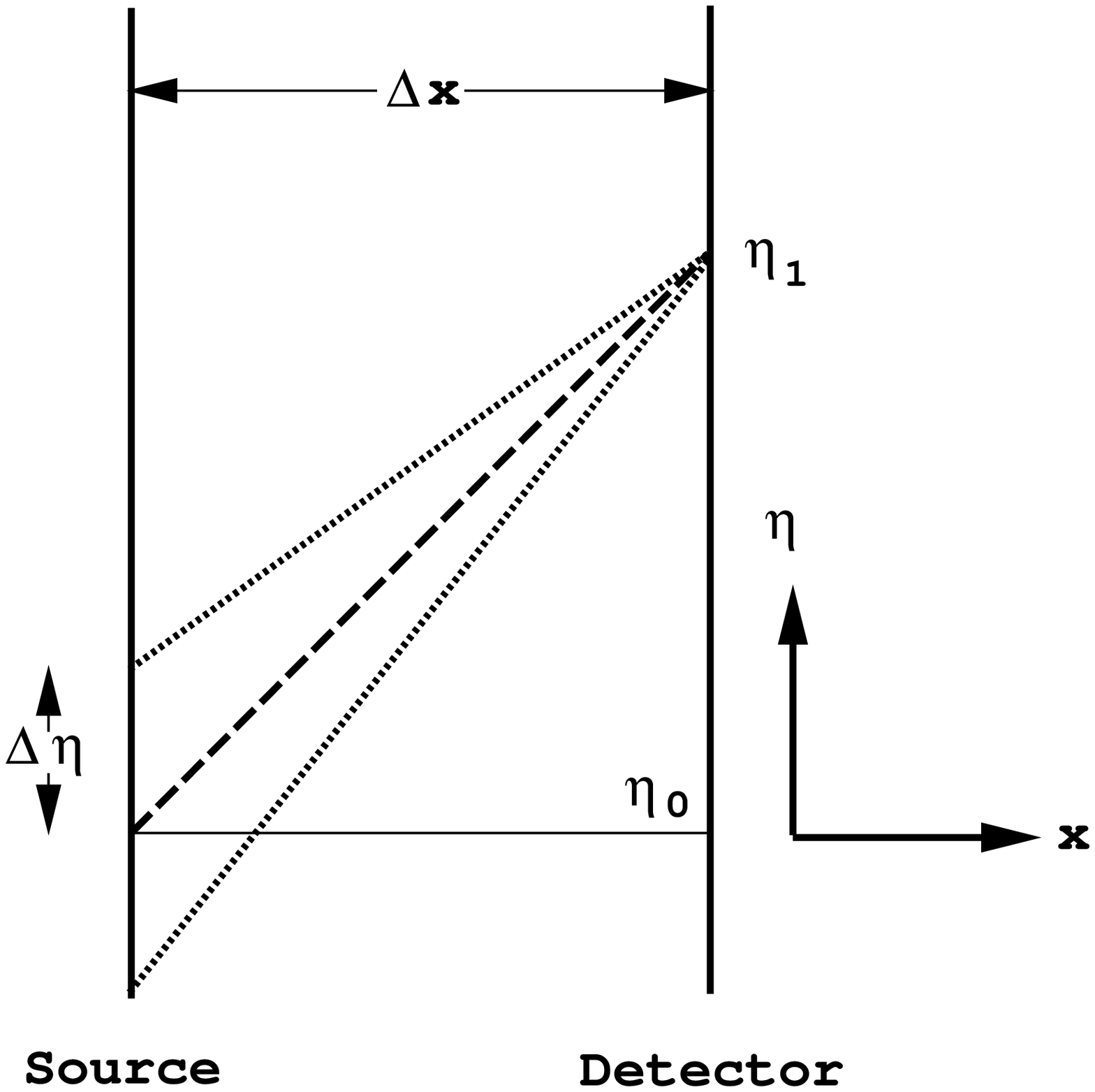}
\label{Figure 1}
\end{center}
\begin{caption}[]

A photon is received by a detector at conformal time $\eta_1$.
In the absence of metric fluctuations, it has traveled along the classical
lightcone (dashed line) from a source at a coordinate distance 
$|\Delta {\bf x}| = \eta_1 -\eta_0$, and was emitted at conformal time $\eta_0$.
In the presence of metric fluctuations, it could have been emitted a
characteristic time $\Delta \eta$ before or after $\eta_0$, and traveled
along a mean trajectory which is either timelike or spacelike, respectively
(dotted lines).
\end{caption}
\end{figure}

    We may interpret this formula by noting that in an expanding universe, 
$\beta^{-1}$ is a coordinate temperature, not in general the physical 
temperature. It is, however, the physical temperature at a time at which
$a = 1$. Let us take that time to be $t_0$, the time of emission, at which 
time the physical temperature is $T_0$. This leads to
\begin{equation}
\frac{\Delta t}{t_0} = \frac{\sqrt{6}}{6} \sqrt{\ln(\eta_0/\epsilon) + 1}
  \, \biggl(\frac{T_0}{T_p}\biggr) \biggl(\frac{t_p}{t_0}\biggr) \,,
\end{equation}
where $T_p$ is the Planck temperature, and $t_p$ is the Planck time.
The logarithmic factor can be taken to be of order one, so we see that
if $T_0 =T_p$ and $t_0 =t_p$, then we have $\Delta t/t_0 \approx 1$.
Otherwise, with $T_0 \ll T_p$ and $t_0 \geq t_p$, we have $\Delta t/t_0 \ll 1$,
and the fractional lightcone fluctuations are small. This is what we should 
perhaps expect; a bath of gravitons with the Planck temperature at the 
Planck time (which would correspond to a few degrees Kelvin today) results 
in large horizon fluctuations, but otherwise the fluctuations are small
at much lower temperatures.

\subsection{Gravitons in de Sitter Space}
\label{sec:deSit}

    If we represent de Sitter space as a spatially flat Robertson-Walker
metric, then the metric is of the form of Eq.~(\ref{eq:RWmetric}) with
$a(\eta) = -H/\eta = H/|\eta|$, where $H$ is a constant, and 
$-\infty < \eta < 0$. These coordinates
cover one-half of the full de Sitter spacetime, but that is sufficient for our
purposes. Gravitons are again represented as a pair of massless, minimally
coupled scalar fields. However, in this case there is a subtlety in that
there is no de Sitter invariant vacuum state which is free of infrared
divergences \cite{FP77b,VF82,FV86}. The Hadamard function may be represented 
as \cite{FV86}
\begin{equation}
G(x,x') = \frac{1}{(2\pi)^3} {\rm Re} \int d^3{\bf k}\, 
  \psi_k(\eta) \,\psi_k^*(\eta')\, e^{i({\bf k} -{\bf k}')\cdot {\bf x}}\,,
                                               \label{eq:deSitGF}
\end{equation}
where the time part of the mode function is expressible in terms of Hankel
functions as
\begin{equation}
\psi_k(\eta) = \frac{\sqrt{\pi}}{2}\,H\,|\eta|^\frac{3}{2}
 \Bigl[c_1 H_\frac{3}{2}^{(1)}(k\eta) +c_2 H_\frac{3}{2}^{(2)}(k\eta) \Bigr] 
             \,.  \label{eq:modes}
\end{equation}
Here $c_1$ and $c_2$ are functions of $k$ which are required to have the 
following properties:
\begin{equation}
c_1 \rightarrow 0, \quad c_2 \rightarrow 1, \qquad {\rm as}\quad k \rightarrow 
     \infty  \,,        \label{eq:cond1}
\end{equation}
and 
\begin{equation}
|c_1 + c_2| \rightarrow 0,  \qquad {\rm as}\quad k \rightarrow 0 \,.
                                                 \label{eq:cond2}
\end{equation}
These functions define the quantum state in question. Equation~(\ref{eq:cond1})
insures that the very high frequency modes are free of particles, whereas
Eq.~(\ref{eq:cond2}) is the condition that the state be free of infrared 
divergences.

   The Hadamard function given by Eqs.~(\ref{eq:deSitGF}) and (\ref{eq:modes})
is the unrenormalized function which is singular on the lightcone. In
principle, one should evaluate the integral for the given choice of $c_1$
and $c_2$, and then extract the state-independent singular terms to obtain
the renormalized Hadamard function. Unfortunately, this would be very difficult
to do explicitly. However, in the late time limits ($\eta \rightarrow 0$ or
$\eta' \rightarrow 0$) it is possible to give some approximate forms for 
the renormalized function. It was shown by several authors that the coincidence 
limit grows logarithmically \cite{VF82,FV86,L82,S82} :
\begin{equation}
G(x,x) \sim - \frac{H^2}{4 \pi^2}\, \ln H|\eta| \,,
                                                    \label{eq:deSitGF4}
\end{equation}
as $\eta =\eta' \rightarrow 0$.  The fact that
this asymptotic form is state-independent may be understood as a consequence
of the exponential expansion having redshifted away any memory of the quantum
state. 

 We may use a similar
procedure to investigate the behavior as $\eta  \rightarrow 0$ with $\eta'$
fixed. If we insert Eq.~(\ref{eq:modes}) into Eq.~(\ref{eq:deSitGF}) and then
change the variable of integration to ${\bf q} = |\eta-\eta'|\, {\bf k}$,
 we have
\begin{eqnarray}
G(x,x') &=& \frac{H^2 |\eta\eta'|^{\frac{3}{2}} }{32\pi^2\,|\eta-\eta'|^3}\;
{\rm Re} \int d^3{\bf q}\, 
\biggl[ c_1 H_\frac{3}{2}^{(1)}\biggl(-\frac{|\eta|}{|\eta-\eta'|}\, q \biggr)
+ c_2 H_\frac{3}{2}^{(2)}\biggl(-\frac{|\eta|}{|\eta-\eta'|}\, q \biggr)\biggr]
                            \nonumber \\
&\times&
\biggl[c_1^* H_\frac{3}{2}^{(2)}\biggl(-\frac{|\eta'|}{|\eta-\eta'|}\, q \biggr)
+c_2^* H_\frac{3}{2}^{(1)}\biggl(-\frac{|\eta'|}{|\eta-\eta'|}\, q \biggr)\biggr]
\; e^{i{\bf v} \cdot {\bf q}}  \,, \label{eq:deSitGF2}
\end{eqnarray}
where ${\bf v} = ({\bf x} - {\bf x'})|\eta-\eta'|^{-1}$. Now assume that
$\eta $ is sufficiently small that the dominant contribution
to the integral comes from values of ${\bf q}$ for which the magnitude of the
arguments of the Hankel functions in the first factor  are small compared to 
unity. In this case, we can use the small argument forms
\begin{equation}
H_\frac{3}{2}^{(1)}(-x) \approx H_\frac{3}{2}^{(2)}(-x) \approx 
\sqrt{\frac{2}{\pi}} \, x^{-\frac{3}{2}} \,, \qquad 0<x\ll 1,
\end{equation}
in this factor and find 
\begin{equation}
G(x,x') \sim \frac{\sqrt{2}\, H^2}{32 \pi^{5/2}}\; {\rm Re} 
\int d^3{\bf q}\, q^{-\frac{3}{2}}\, (c_1 + c_2)
\bigl[c_1^* H_\frac{3}{2}^{(2)}(-q) +c_2^* H_\frac{3}{2}^{(1)}(-q)\bigr]
          \, e^{i{\bf v} \cdot {\bf q}}  \,. \label{eq:deSitGF3}
\end{equation}
Note that the dependence upon both $\eta$ and $\eta'$ has dropped out. 
The integral in Eq.~(\ref{eq:deSitGF3}) has a logarithmic ultraviolet 
divergence on the lightcone, but the leading quadratic divergence has 
disappeared. A similar disappearance of divergent parts occurs in the
derivation of Eq.~(\ref{eq:deSitGF4}). (See, for example, Ref. \cite{FV86}.)
The renormalization of this logarithmic divergence requires a subtraction of 
a term proportional to the scalar curvature $R = 12 H^2$. We expect the result
to be a constant which is of order of $H^2$. By symmetry, we obtain the same 
result if  $\eta'  \rightarrow 0$ with $\eta$ fixed, so that
\begin{equation}
G(x,x') \approx H^2\, ,
   \qquad \eta \rightarrow 0\;\; {\rm or}\;\; \eta'  \rightarrow 0\,.
                                                \label{eq:deSitGF5}
\end{equation}
The integral in Eq.~(\ref{eq:sig1}) requires us to know $G(x,x')$ in the
square illustrated in Fig. 2, which also illustrates the regions in which
either Eq.~(\ref{eq:deSitGF4}) or Eq.~(\ref{eq:deSitGF5}) is applicable.
For the purposes of obtaining an order of magnitude estimate for 
$\langle \sigma_1^2 \rangle$, we will assume that $G(x,x')$ is of the form
\begin{equation}
G(x,x') \approx H^2\, F(\eta,\eta')
\end{equation}
throughout this region, where $F$ is either a constant of order unity or else
a logarithmic function which will contribute multiplicative constants 
of order unity to the integral in Eq.~(\ref{eq:sig1}).

\begin{figure}
\begin{center}
\leavevmode\epsfysize=10cm\epsffile{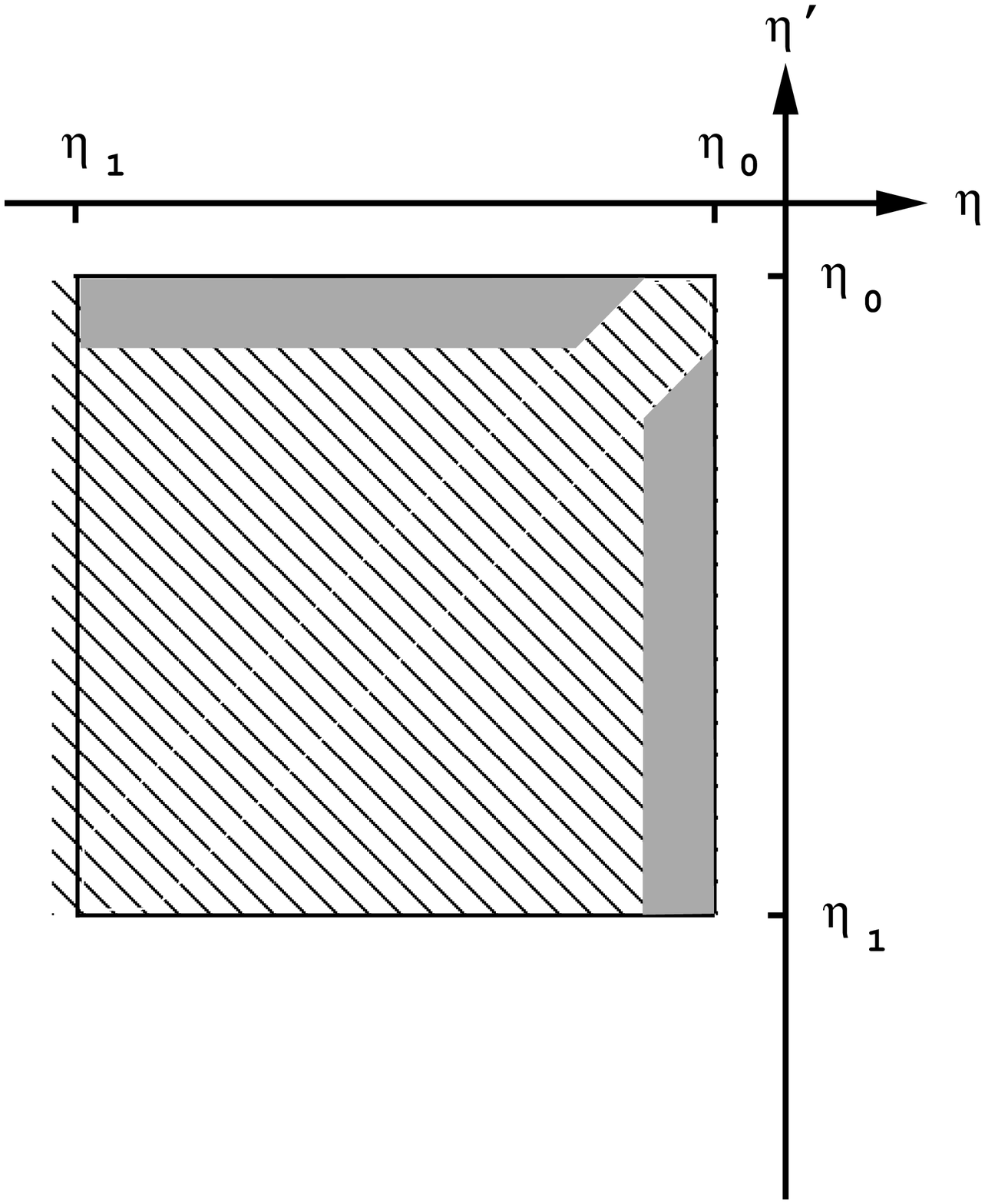}
\label{Figure 2}
\end{center}
\begin{caption}[]

The domain of integration for $\langle \sigma_1^2 \rangle$ is the
interior of the square. The integrand, $G(x,x')$ is known in the shaded
regions to be approximately $H^2$, and in the crosshatched region to be
approximately $- ({H^2}\, \ln H|\eta|)/{4 \pi^2}$.
\end{caption}
\end{figure}

 The result of the evaluation of this integral is
\begin{equation}
\langle \sigma_1^2 \rangle = \kappa\, \frac{H^2 \sigma_0}{\alpha}
                     (\eta_1 -\eta_0)^2  \,,
\end{equation}
where $\kappa$ is a constant of order unity.  From Eq~(\ref{eq:s0}) we find
\begin{equation}
\sigma_0 = \frac{\alpha (\eta_1 -\eta_0)^2}{H^4\, \eta_1^2\, \eta_0^2} \,.
\end{equation}
The  value of $\alpha$ which is characteristic of the lightcone fluctuations
is
\begin{equation}
\alpha_c \approx  \sqrt{2\kappa}\, H^3\, \eta_1 \eta_0\, 
            \,.
\end{equation}
If we follow the line of reasoning in the previous subsection, we find
that the characteristic time associated with the de Sitter horizon fluctuations
is 
\begin{equation}
\Delta t \approx  \sqrt{2\kappa}\,.
\end{equation}
This is of order unity, and hence again the horizon fluctuations are of 
Planck dimensions.

    We may use $\Delta t$ to find the frequency fluctuations observed at
$\eta_1$ from a constant frequency source. Suppose that the source is emitting
photons with a constant frequency $\nu_0$. In the absence of metric fluctuations,
the photons will be detected at frequency $\nu= \nu_0\, a(\eta_0)/a(\eta_1)$.
The effect of the metric fluctuations is equivalent to a drift in the source
frequency whose magnitude is $\Delta \nu_0 = \nu_0^2\, \Delta t$. Consequently,
the fractional variation of frequency at the detector is
\begin{equation}
\frac{ \Delta \nu}{\nu} = \nu_0\, \Delta t\, . \label{eq:freqvar}
\end{equation}
However, this is an ensemble averaged frequency variation, not necessarily 
the drift in frequency that would be observed in any one trial. The reason 
for this is that pulses emitted close together in time tend to have correlated
time delays or advances \cite{FS96}.  Thus, Eq.~(\ref{eq:freqvar}) should be
interpreted as giving an upper bound in the frequency drift seen by the 
detector.

\section{Black Hole Horizons}
\label{sec:BH}

    Here we wish to discuss the fluctuations of the event horizon of a
Schwarzschild black hole, for which the metric is
\begin{equation}
ds^2 = \biggl(1 -\frac{2M}{r}\biggr) dt^2 - 
 \biggl(1 -\frac{2M}{r}\biggr)^{-1}dr^2 -r^2(d\theta^2 +\sin^2\theta \, d\phi^2)
                                          \,.
\end{equation}
Timelike radial geodesics in this metric satisfy \cite{MTW}
\begin{equation}
\biggl(\frac{dr}{d\tau}\biggr)^2 = \tilde E^2 - C(r)  \label{eq:drdtau}
\end{equation}
and 
\begin{equation}
\frac{dt}{d\tau} = \tilde E/C \,, \label{eq:dtdtau}
\end{equation}
where $C(r) = 1-2M/r$ and $\tilde E$ is a constant of the motion which is
equal to the energy per unit rest mass of the particle, as measured at infinity.
From these relations, one may show that
\begin{equation}
\biggl(\frac{dr}{d t}\biggr)^2 = C^2 \biggl(1 - \frac{C}{\tilde E^2}\biggr)\,,
                                                     \label{eq:drdt}   
\end{equation}
and that 
\begin{equation}
\biggl(\frac{dr^*}{d t}\biggr)^2 =1 - \frac{C}{\tilde E^2}\,, \label{eq:dr*dt}
\end{equation}
where 
\begin{equation}
\frac{d r}{d r^*} = C\,.
\end{equation}
From Eq.~(\ref{eq:dtdtau}), we see that the proper time elapsed along a 
segment of a geodesic is 
\begin{equation}
\Delta \tau = {\tilde E}^{-1} \int_{t_0}^{t_1} C(r) dt \,,
\end{equation}
where $r$ is understood to be a function of $t$ along the geodesic.
We are primarily interested in the case of nearly null outgoing geodesics, for 
which $\tilde E = \tilde E_o \gg 1$. In this limit
\begin{equation}
\frac{dr}{d t} \approx  C(r) \,.
\end{equation}
 Thus, $|\Delta \tau| \approx |\Delta r|/\tilde E_o$ and
\begin{equation}
\sigma_0 \approx \frac{(\Delta r)^2}{2 \tilde E_o^2} \,,  \label{eq:sig0BH}
\end{equation}
where $\Delta r$ is the radial coordinate interval traversed. An analogous
treatment may be given for spacelike geodesics. The constant of the motion
$\tilde E$ no longer has a simple physical interpretation, but we can express
$\sigma_0$ in this case as
\begin{equation}
\sigma_0 \approx - \frac{(\Delta r)^2}{2 \tilde E_o^2} \,.
\end{equation}

   We now turn to the task of estimating $\langle \sigma_1^2 \rangle$
near the horizon of a Schwarzschild black hole. From Eq.~(\ref{eq:sig1gen})
and the fact that
\begin{equation}
\frac{dr}{d\tau} \approx \tilde E_o
\end{equation}
for nearly null outgoing timelike geodesics, we have that
\begin{equation}
\langle \sigma_1^2 \rangle \approx \frac{1}{2} \sigma_0 \tilde E_o^{-2}
\int dr_1\, dr_2\, u_1^\mu u_1^\nu u_2^\rho u_2^\sigma \,
                     \langle h_{\mu\nu}(x_1) h_{\rho\sigma}(x_2) \rangle \,.
                                 \label{eq:sig1BH}
\end{equation}
If we are interested in a black hole radiating into empty space, then the 
relevant quantum state for the quantized graviton field is the Unruh state.
It would be a rather formidable task to explicitly compute the renormalized
graviton two point function in the state. Instead, we will content ourselves
with an order of magnitude estimate. First we must choose a convenient gauge.
Again we wish to impose the transverse, tracefree gauge, which eliminates all
gauge freedom. Because all of the modes of the graviton field are propagating
waves which either originate at ${\cal I}^-$ or reach ${\cal I}^+$, we can
impose the requirement that these modes satisfy the flat space transverse, 
tracefree gauge condition at $r = \infty$. 

   We now make the assumption that in this gauge, the renormalized two point
function measured in the frame of an infalling observer who starts from infinity
at rest can be estimated by dimensional considerations. Near the horizon at
$r = 2M$, the geometry and the quantum state are characterized by a single
scale, $M$. If we were to reinstate explicit factors of Newton's constant,
$G$, then $h_{\mu\nu} \propto G^{-\frac{1}{2}} \propto m_p$ where $m_p$ is 
the Planck mass. However, $h_{\mu\nu}$ is dimensionless in any set of units,
so our assumption tells us that the two point function should be proportional
to $m_p^2/M^2$. However, the actual values of the components of this bitensor
depend upon the choice of frame. Our assumption is that infalling observers
with $\tilde E = \tilde E_i \approx 1$ should be regarded as preferred in 
the sense that they 
do not introduce any very large or very small dimensionless redshift or
blueshift factors. Let $v^\mu$ be the four velocity of such an observer.
Our assumption may be expressed as
\begin{equation}
v_1^\mu v_1^\nu v_2^\rho v_2^\sigma \,  
\langle h_{\mu\nu}(x_1) h_{\rho\sigma}(x_2) \rangle \approx \frac{m_p^2}{M^2}
                                                \label{eq:hsq1}
\end{equation}
near the horizon. The components of the infalling observer's four velocity are
\begin{equation}
v^t = C^{-1}, \qquad v^r = -\sqrt{1 - C} \approx -1 \,,
\end{equation}
and those for an outgoing observer are
\begin{equation}
u^t = \tilde E_o \,C^{-1}, \qquad u^r = \sqrt{\tilde E_o^2 - C} 
        \approx \tilde E_o \,.
\end{equation}
Thus $|u^\mu| = \tilde E_o \,|v^\mu|$ and we can write
\begin{equation}
u_1^\mu u_1^\nu u_2^\rho u_2^\sigma \,  
\langle h_{\mu\nu}(x_1) h_{\rho\sigma}(x_2) \rangle 
                     \approx \tilde E_o^4 \,\frac{m_p^2}{M^2} \,.
                                                \label{eq:hsq2}
\end{equation}
Our basic assumption receives some support from the work of York \cite{York}
who estimates the magnitude of the quantum fluctuations of the lowest modes
of vibration of a Schwarzschild black hole. He treats these modes as quantum
mechanical harmonic oscillators, and calculates their root-mean-square
fluctuation amplitudes. The amplitudes of the first few 
modes yields a result consistent with Eqs.~(\ref{eq:hsq1}) or (\ref{eq:hsq2}).
Of course, this is heuristic support, and by no means a proof of our 
assumption. A full proof would require one to sum over an infinite number of
degrees of freedom, and then extract any ultraviolet divergent parts.

    The graviton two point function in this approximation is a constant in the 
vicinity of the horizon. It must also fall off to zero at large distances from
the black hole. Thus the integral in Eq.~(\ref{eq:sig1BH}) gets its dominant
contribution over an interval in $r$ of the order of $M$, regardless of the
upper limit of the integration. In any case, we can stop the integration at 
a maximum value of $r$ which is just a few times $M$. Whether the outgoing
photons emitted in the vicinity of the horizon are detected at $r = 4M$ or
at a much larger value of $r$ has little effect on the discussion of the
horizon fluctuations. Thus we may let
\begin{equation}
\langle \sigma_1^2 \rangle \approx 
\sigma_0 \, \tilde E_o^2 \,\frac{(\Delta r)^2}{M^2} \,.
                                        \label{eq:sig1BH2}            
\end{equation}
In analogy to the discussion in the previous section, we wish to define a
characteristic value $\tilde E_c$, which is the value of $\tilde E_o$ at
which Eq.~(\ref{eq:alphac}) holds. From  Eqs.~(\ref{eq:sig0BH}) and 
(\ref{eq:sig1BH2}) we find
\begin{equation}
\tilde E_c  \approx  \sqrt{M} \,. \label{eq:etildec}
\end{equation}
We may find the associated time delay or advance, $\Delta t$, from 
Eq.~(\ref{eq:dr*dt}), which tells us that when $\tilde E_o \gg 1$
\begin{equation}
dt \approx dr^* + \frac{C}{2 \tilde E_o^2}\, dr^* = 
                       dr^* + \frac{1}{2 \tilde E_o^2}\, dr \,.
\end{equation}
A radial null geodesic in the classical background geometry covers an 
$r^*$ distance of $\Delta r^*$ in a coordinate time $\Delta t =\Delta r^*$. 
The second term on the right hand side of the above equation tells us the
extra amount of time required by a timelike particle. Analogous expressions 
hold for spacelike geodesics, and yield the same magnitude of time variation.
Thus we are led to
an expression for the characteristic time delay or advance due to horizon
fluctuations:
\begin{equation} 
\Delta t \approx \frac{\Delta r}{M} \,.
\end{equation}
As discussed above, we can take $\Delta r$ to be of order $M$, although we
might also want to consider the possibility of taking it to be much smaller.
Thus let
\begin{equation}
\Delta r = \gamma \, M \,,
\end{equation}
where $\gamma$ is a constant of the order of or less than unity.
 Now we have
\begin{equation} 
\Delta t \approx \gamma \,,
\end{equation}
so the time delay, measured in coordinate time, is of Planck dimensions.
However, a more physical measure is obtained by expressing this time interval
in terms of the proper time of a local observer. Let the photons be emitted at
$r=r_0 = 2M(1 +\epsilon)$, with $\epsilon \ll 1$, and let $C_0 = C(r_0)
\approx \epsilon$. The time interval in the frame of a static (nongeodesic)
observer at rest at $r=r_0$ is
\begin{equation}
\Delta \tau_s \approx \gamma\, \sqrt{C_0} \,, 
\end{equation}
that in the frame of an infalling observer with  $\tilde E = 1$ is
\begin{equation}
\Delta \tau_i \approx \gamma \, {C_0} \approx \gamma\, \epsilon \,, 
                                                        \label{eq:taui}
\end{equation}
and that in the frame of an outgoing geodesic observer with $\tilde E = 1$ is
\begin{equation}
\Delta \tau_o \approx \gamma \,. 
\end{equation}

One might regard $\Delta \tau_i$, the characteristic time as measured by an
infalling observer, to be the best measure of the magnitude of the horizon
fluctuations. Such an observer can cross and continue beyond the classical
event horizon at $r= 2M$. Suppose that an outgoing photon emitted by this
observer reaches infinity. An observer at infinity who detects this photon
and who is unaware of the lightcone fluctuations might trace the history of
this photon backwards in the classical Schwarzschild geometry and infer that
it was emitted at a proper time of $\tau_0$ on the infalling observer's 
worldline. In fact, it could have been emitted anywhere in a band of width
$\Delta \tau_i$ centered around $\tau_0$. (See Fig. 3.)
The remarkable feature of the result
Eq.~(\ref{eq:taui}) is that $\Delta \tau_i \rightarrow 0$ as 
$\tau_0 \rightarrow \tau_H$, the proper time at which the infalling observer
reaches $r= 2M$. In the cosmological models discussed in Sect.~\ref{sec:Cosmo},
the fluctuation in emission time was typically of the order of the Planck
time. In the black hole case, the horizon fluctuations are more strongly
suppressed. Note that the proper time required for the infalling observer
to pass from $r=r_0 = 2M(1 +\epsilon)$ to $r= 2M$ is $T \approx 2\epsilon M$.
This is always large compared to $\Delta \tau_i$ for large black holes:
\begin{equation}
\frac{\Delta \tau_i}{T} \approx \gamma \frac{m_p}{M} \,.
\end{equation}
Thus the only outgoing photons which manage to cross the classical horizon
are part of an extreme tail of a Gaussian distribution.

\begin{figure}
\begin{center}
\leavevmode\epsfysize=8cm\epsffile{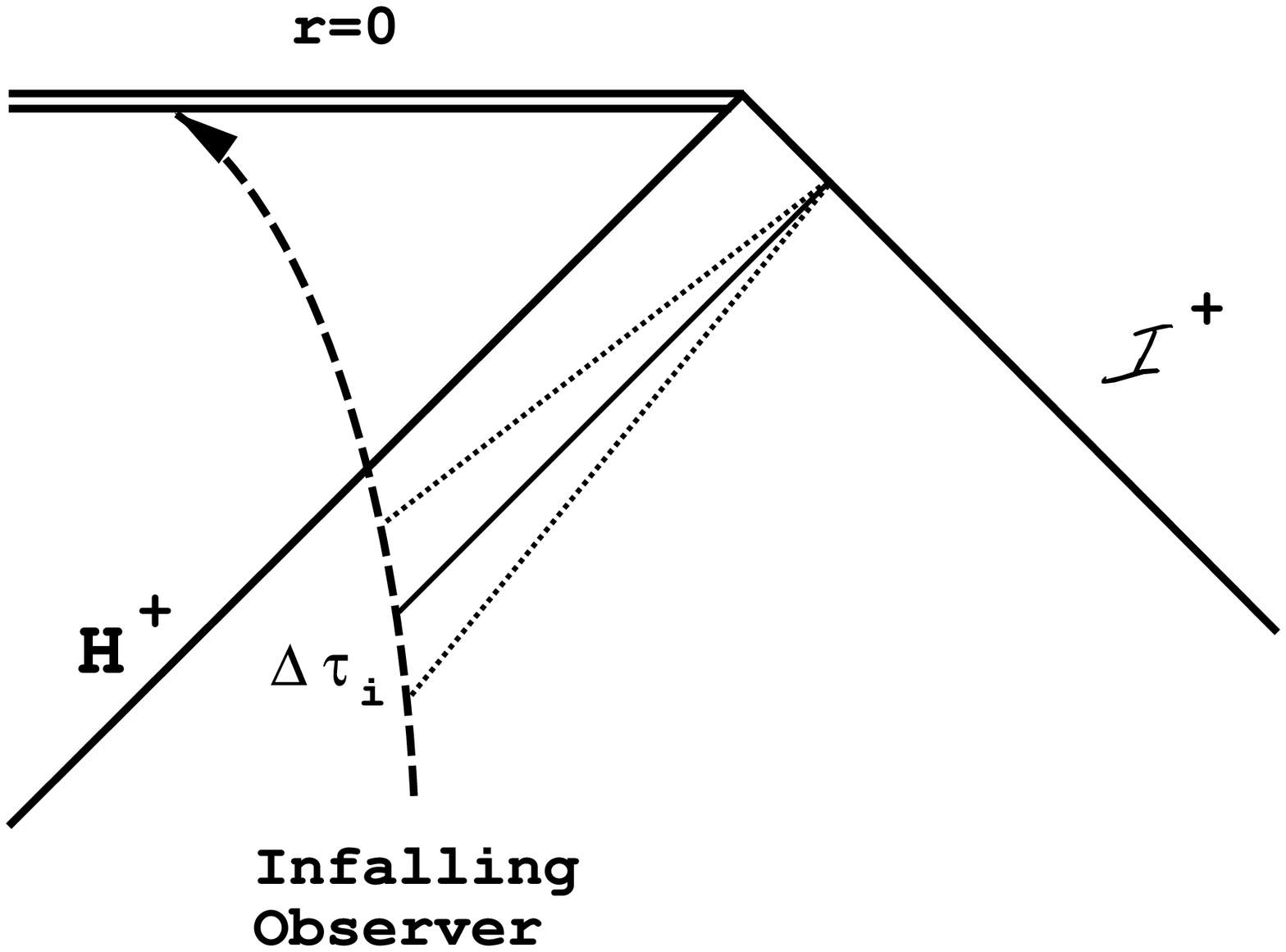}
\label{Figure 3}
\end{center}
\begin{caption}[]

An observer falling across the future horizon, $H^+$, of a black
hole emits photons which reach ${\cal I}^+$. In the presence of metric 
fluctuations, these photons need not follow the classical lightcone (solid
line), but rather may follow timelike or spacelike paths in the background
geometry (dotted lines). The characteristic variation in emission time,
as measured in the frame of the infalling observer, of photons which reach
${\cal I}^+$ at the same point is $\Delta \tau_i$.
\end{caption}
\end{figure}

   As in Sect. \ref{sec:deSit}, we may express the time delay or advance 
in terms of the variation in frequency seen by the observer at infinity.
In the black hole case, the analog of Eq.~(\ref{eq:freqvar}) is
\begin{equation}
\frac{ \Delta \nu}{\nu} = \nu_0\, \Delta \tau_i
       \approx \nu_0\, \gamma\, \epsilon    \, . \label{eq:freqvar2}
\end{equation}
Thus as the source approaches $r=2M$, the fractional variation in frequency
observed at infinity goes to zero, and the observed frequency approaches that
predicted by classical relativity.

  Let us now turn to the question of whether horizon fluctuations are
capable of invalidating the semiclassical derivation of the Hawking effect.
First let us recall the essential features of this derivation, as given in
Hawking's original paper \cite{Hawking75}. Consider the spacetime of a
black hole formed by gravitational collapse (Figure 4). The null ray which
forms the future horizon leaves ${\cal I}^-$ at advanced time $v = v_0$.
The modes into which the outgoing thermal radiation will be created leave
${\cal I}^-$ at values of $v$ slightly less than $v_0$, pass through the
collapsing body, and reach ${\cal I}^+$ as outgoing rays, on which the
retarded time $u$ is constant. Hawking shows that the relation between the
values of $v$ and of $u$ is
\begin{equation}
u = -4M\, \ln\biggl( \frac{v_0 -v}{A} \biggr) \,,
\end{equation}
where $A$ is a constant. Thus $u \rightarrow \infty$ as $v \rightarrow v_0$.
As seen by an observer at infinity, these outgoing rays must hover extremely
close to the horizon for a very long time. If one starts with a black hole
with a mass $M$ large compared to the Planck mass, the semiclassical
description should hold for the time required for the black hole to lose most of 
its original mass. Let
\begin{equation}
t_{evap} = M^3 = M \biggl(\frac{M}{m_p}\biggr)^2
\end{equation}
be this characteristic evaporation time. The basic problem posed by the horizon
fluctuations is that they may cause an outgoing ray to either fall back into
the black hole, or else to prematurely escape. In either case, the 
semiclassical picture of black hole radiance would need to be modified at
times less than $t_{evap}$. 

\begin{figure}
\begin{center}
\leavevmode\epsfysize=10cm\epsffile{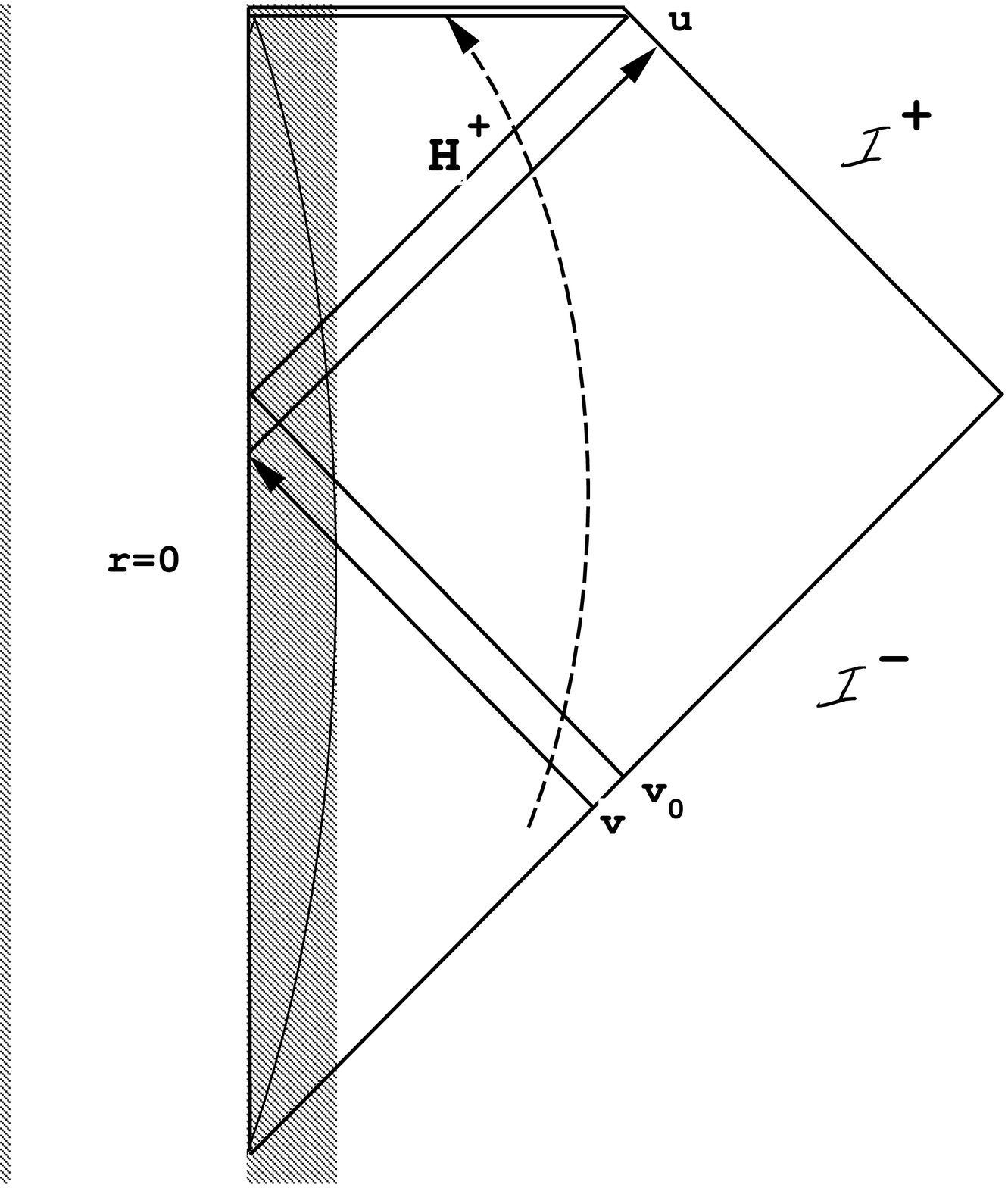}
\label{Figure 4}
\end{center}
\begin{caption}[]

The spacetime for a black hole formed by gravitational collapse.
The shaded region is the interior of the collapsing star. A null ray which
leaves ${\cal I}^-$ with advanced time $v_0$ becomes the future horizon,
$H^+$. A ray which leaves at an earlier time $v$ passes through the
collapsing body and reaches ${\cal I}^+$ at retarded time $u$. The dashed
line is the worldline of a observer who falls into the black hole after its
formation.
\end{caption}
\end{figure}

  At a large distance from the black hole, $u=t -r^* \approx t -r$. If
the observer at ``infinity'' is at a fixed value of $r$ (e.g. $100M$), then
$u \approx t$ for most of the black hole's lifetime. Thus, in order not
to invalidate the semiclassical treatment, outgoing rays with $u < u_{max}
=t_{evap}$ need to be uninfluenced by the horizon fluctuations. 
In order to investigate this question, let us consider an infalling observer 
with $\tilde E =\tilde E_i =1$. From  Eq.~(\ref{eq:drdt}), we have that 
near $r=2M$
\begin{equation}
\frac{dt}{d r} \approx - C^{-1}
\end{equation}
and hence
\begin{equation}
\frac{du}{d r} = \frac{dt}{d r} - \frac{dr^*}{d r} \approx - 2C^{-1}\,.
\end{equation}
This equation may be integrated to yield
\begin{equation}
u(r) = -4M\, \ln\biggl( \frac{r - 2M}{A'} \biggr) \,,
\end{equation}
where $A'$ is a constant. This relation tells us the value of $r$ at which
the infalling observer crosses a given constant $u$ line. The constant $A'$
is determined by which infalling observer we consider. Here we are interested    
in observers who fall into the black hole not long after its formation,
and we can set $A' \approx M$. Let $r_c$ be the value of $r$ at which this
observer crosses the $u = u_{max}$ line, given by
\begin{equation}
r_c - 2M = M\, e^{-{u_{max}}/{4M}}\,.
\end{equation}
Near the horizon, Eq.~(\ref{eq:drdtau}) tells us that $dr/d\tau \approx -1$
along the worldline of the infalling observer. Thus the proper time required for 
this observer to cross from $u = u_{max}$ to the classical horizon at $r=2M$
is
\begin{equation}
\delta \tau \approx r_c -2M = M\, e^{-{u_{max}}/{4M}} \approx
                    M\, e^{-{M^2}/{m_p^2}}\,.
\end{equation}
We should compare this quantity with $\Delta \tau_i$, where $C_0$ is evaluated 
at $r=r_c$, so $C_0 = \frac{1}{2} e^{-u_{max}/4M}$. Thus,
\begin{equation}
\Delta \tau_i =  \frac{\gamma m_p}{2M}\, \delta \tau \,,
\end{equation}
and hence so long as $M \gg m_p$, $\Delta \tau_i \ll  \delta \tau$. From this 
result, we conclude that the horizon fluctuations do not invalidate the
semiclassical derivation of the Hawking effect until the black hole's mass
approaches the Planck mass. This is the point at which we would expect the
semiclassical treatment to fail.

The presence of frequencies far above the Planck scale, in the form of the 
modes leaving ${\cal I}^-$, has concerned numerous authors. There have been
suggestions that one might be able derive the Hawking effect in a way that 
transplanckian frequencies do not arise, using some form of ``mode
regeneration'' \cite{Jacobson,Unruh}. So far, it has not been possible to 
implement these suggestions in detail. As seen from the our analysis of
horizon fluctuations, the semiclassical treatment is remarkably robust.

\section{Summary and Conclusions}
\label{sec:Sum}

   In the preceeding sections, we have analyzed the horizon fluctuation
problem using a formalism which takes account of the effects of quantized
linear perturbations of the gravitational field upon lightcones. In the case 
of the cosmological models treated in Sect.~\ref{sec:Cosmo}, the resulting
horizon fluctuations were found to be of Planck dimensions for both de Sitter
space and a radiation filled universe with a Planck density of gravitons at
the Planck time. These fluctuations are measured as fluctuations in the time of 
emission of a photon as measured in the frame of a comoving observer. The
order of magnitude of the results is what one might have guessed before
doing the calculation. 

In the case of black hole horizon fluctuations, the results are somewhat
more subtle. Whether the time scale which characterizes the horizon 
fluctuations (the time delay or advance) is of Planck dimensions or not
depends crucially upon the frame of reference. It is indeed of Planck 
dimensions as measured by an observer at infinity. However, as measured by
an infalling observer, this time is much less than the Planck scale, and
vanishes as the infalling observer approaches the classical event horizon
at $r = 2M$. We further found that this suppression of the horizon fluctuations
is exactly what is needed to preserve Hawking's semiclassical derivation
of black hole radiance for black holes of mass large compared to the Planck
mass. 

Our result seems to conflict with the arguments of Sorkin \cite{Sorkin}
and of Casher {\it et al} \cite{Casher}. These authors claim that the
horizon fluctuations are much larger than found in the present manuscript.
It should be noted, however, that the physical mechanisms being postulated 
in Refs. \cite{Sorkin} and \cite{Casher} are quite different from 
that of the present paper.  Furthermore, in our opinion,
the physical basis of both of these calculations seems to be open to question.
Casher {\it et al} obtain large
gravitational perturbations of the horizon by postulating an ``atmosphere''
of particles near the horizon in large angular momentum modes. This arises 
by decomposing the physical quantum state of an evaporating black hole 
(the Unruh vacuum) into two pieces which  separately have divergent stress
tensors on the horizon, the contribution from the Boulware vacuum state
and a term which these authors call the ``atmosphere'' of particles. The 
large stress tensor fluctuations arise in the Casher {\it et al} analysis
when this ``atmosphere'' undergoes thermal fluctuations. Our objection to this
procedure is that the fluctuations of the Boulware vacuum energy density are
not being considered. The splitting of the finite Unruh vacuum energy 
density into two singular parts seems rather artificial. If one chooses such a
splitting, then care must be taken to prove that fluctuations in one part are 
not cancelled by correlated fluctuations in the other part. Casher {\it et al}
have not done this.

Sorkin \cite{Sorkin} uses a Newtonian treatment to
estimate the gravitational field of a mass fluctuation near the horizon, and 
its effects on the Schwarzschild geometry. One can certainly question whether
a Newtonian analysis can be trusted in black hole physics. However, our
primary objection to Sorkin's treatment is that the dominant contribution
to the horizon fluctuations comes from modes whose wavelength is very small
compared to the size of the black hole. The same line of reasoning would seem 
to lead to large stress tensor fluctuations, and hence large lightcone 
fluctuations, in all spacetimes including flat spacetime. In our view,
a more reasonable result is one in which significant fluctuations arise
only on scales characterized either by the spacetime geometry, or else the
chosen quantum state. An approach to defining stress tensor fluctuations
on a flat background which has this property was given in Ref. \cite{KF93}.
Here the stress tensor fluctuations are defined in terms of products of
operators which are normal ordered with respect to the Minkowski vacuum state.

Recently, the fluctuations of the Hawking flux, as measured in the asymptotic
region, have been computed \cite{WF97} by a similar approach. It was found
that this flux undergoes fluctuations of the same order as its average value
over time scales of the order of $M$. This average flux is of order $M^{-2}$,
so the characteristic associated black hole mass fluctuation
is of order $M^{-1}$. The corresponding metric fluctuation near the horizon
is then of order $\delta h \approx M^{-2}$. For macroscopic black holes, this
is much smaller than the metric fluctuations due to the quantized linear
perturbation, estimated in Eq.~(\ref{eq:hsq1}) to be of order $M^{-1}$.
This analysis does not rule out the possibility of much larger stress tensor
fluctuations in the vacuum energy near the horizon. However, the diagonal
and off-diagonal components of the expectation value of the stress tensor
in the Unruh state near the horizon are of the same order \cite{JMO91}.
It is thus plausible that the fluctuations in these various components near 
the horizon are also of the same order. If so, then the effects of quantized
linear perturbations of the gravitational field dominate over those of 
stress tensor fluctuations.

It must be emphasized that all of the conclusions obtained in the present
manuscript are in the context of a model of linearized quantum gravity.
Furthermore, much of our discussion is of a heuristic, order-of-magnitude
nature. If the basic picture of horizon fluctuations which we have drawn is
correct, much work remains to be done to make the picture more precise.

\vspace {0.5 in}
{\bf Acknowledgement:} We would like to thank Tom Roman, Alex Vilenkin and 
Serge Winitzky
for helpful discussions. This work was supported in part by the National
Science Foundation under Grant PHY-9507351 and by Conselho Nacional de
Desevolvimento Cientifico e Tecnol{\'o}gico do Brasil (CNPq).


\begin{thebibliography}{--}

\bibitem{BM95} J.D. Bekenstein and  V. F. Mukhanov,  Phys. Lett. {\bf B360},
7 (1995), gr-qc/9505012.

\bibitem{Sorkin} R.D. Sorkin, {\it Two Topics concerning Black Holes: 
Extremality of the
Energy, Fractality of the Horizon}, gr-qc/9508002; {\it How Wrinkled is the
Surface of a Black Hole?}, gr-qc/9701056.

\bibitem{Casher} A. Casher, F. Englert, N. Itzhaki, and R. Parentani,
{\it Black Hole Horizon Fluctuations}, hep-th/9606106.

\bibitem{F95} L.H. Ford, Phys. Rev. D {\bf 51}, 1692 (1995),  gr-qc/9410047.

\bibitem{FS96} L.H. Ford and N.F. Svaiter,  Phys. Rev. D {\bf 54}, 2640 (1996),
 gr-qc/9604052.

\bibitem{units} Units in which $\hbar = c = 16 \pi G =1$ will be used in this
paper. Thus the units of mass, length and time differ by factors of
$\sqrt{16 \pi}$ from the usual definitions of the Planck mass, Planck length,
and Planck time. As most of the discussion in this paper deals with order
of magnitude estimates, this should not cause any confusion. 
The metric signature will be $(1,-1,-1,-1)$.

\bibitem{BO86} M.R. Brown and A.C. Ottewill, Phys. Rev. D {\bf 34}, 1776 (1986).

\bibitem{AMO92} B. Allen, J.G. McLaughlin, and A.C. Ottewill, Phys. Rev. D 
{\bf 45}, 4486 (1992). 

\bibitem{FP77} L.H. Ford and L. Parker, Phys. Rev. D {\bf 16}, 1601 (1977).

\bibitem{FP77b} L.H. Ford and L. Parker, Phys. Rev. D {\bf 16}, 245 (1977).

\bibitem{VF82} A. Vilenkin and L.H. Ford, Phys. Rev. D {\bf 26}, 1231 (1982).

\bibitem{FV86} L.H. Ford and A. Vilenkin, Phys. Rev. D {\bf 33}, 2833 (1986).

\bibitem{L82} A. Linde, Phys. Lett. {\bf 116B}, 335 (1982).

\bibitem{S82} A.A. Starobinsky, Phys. Lett. {\bf 117B}, 175 (1982).

\bibitem{York} J.W. York, Phys. Rev. D {\bf 28}, 2929 (1983).

\bibitem{MTW} C.W. Misner, K. Thorne, and J.A. Wheeler, {\it Gravitation}
(Freeman, San Francisco, 1973), Sect. 25.5.

\bibitem{Hawking75} S.W. Hawking, Comm. Math, Phys. {\bf 43}, 199 (1975).

\bibitem{Jacobson} T. Jacobson, Phys. Rev. D {\bf 48}, 728 (1993), 
 hep-th/9303103; {\bf 53}, 7028 (1996),  hep-th/9601064.

\bibitem{Unruh} W. G. Unruh, {\it Dumb Holes and the Effects of High 
Frequencies on Black Hole Evaporation},  gr-qc/9409008.

\bibitem{KF93} C.-I Kuo and L.H. Ford, Phys. Rev. D {\bf 47}, 4510 (1993),
 gr-qc/9304008.

\bibitem{WF97} C.H. Wu and L.H. Ford, manuscript in preparation.

\bibitem{JMO91} B.P. Jensen, J.G. McLaughlin, and A.C. Ottewill, Phys. Rev.
              D {\bf 43}, 4142 (1991).

\end{thebibliography}
\end{document}